\documentclass{optica-article}

\journal{opticajournal} % for journals or Optica Open

\articletype{Research Article}

\usepackage{lineno}
\usepackage{bbm}
\usepackage{mathrsfs}
\usepackage{amsmath}

\begin{document}

\title{Generation of multipartite entangled states based on double-longitudinal-mode cavity optomechanial system}

\author{Xiaomin Liu,\authormark{1,2} RongGuo Yang,\authormark{1,2,4} Jing Zhang,\authormark{1,2,4,*} and Tiancai Zhang\authormark{1,3,4}}

\address{\authormark{1}College of Physics and Electronic Engineering, Shanxi University, Taiyuan 030006, China\\
\authormark{2}State Key Laboratory of Quantum Optics and Quantum Optics Devices, Shanxi University, Taiyuan 030006, China\\
\authormark{3}Institute of Opto-Electronics, Shanxi University, Taiyuan 030006, China\\
\authormark{4}Collaborative Innovation Center of Extreme Optics, Shanxi University, Taiyuan 030006, China}
\email{\authormark{*}zjj@sxu.edu.cn} %% email address is required; see note below about the corresponding author designation
% use {asbstract*} to suppress the copyright line. Copyright information will be added in production

\begin{abstract*} 
Optomechanical system is a promising platform to connect different “notes” of quantum networks, therefore, entanglement generated from it is also of great importance. In this paper, the parameter dependence of optomechanical and optical-optical entanglements generated from the double-longitudinal-mode cavity optomechanical system are discussed and two quadrapartite entanglement generation schemes based on such a system are proposed. Furthermore, 2N or 4N-partite entangled states can be obtained by coupling N cavities with N-1 beamsplitter(BS)s, and these schemes are scalable in increasing the partite number of entanglement. Certain ladder or linear structures are contained in the finally obtained entanglement structure, which can be applied in quantum computing or quantum networks in the future. 
\end{abstract*}

%%%%%%%%%%%%%%%%%%%%%%%%%%  body  %%%%%%%%%%%%%%%%%%%%%%%%%%
\section{Introduction}
Quantum entanglement, as a profound feature of quantum theory, is the prerequisite of quantum teleportation\cite{169}, quantum dense coding\cite{Guo2019}, quantum computing\cite{45}, quantum communication\cite{965}, and quantum networks\cite{01226}. With the development of quantum information technology, multipartite entanglement is required for realizing measurement-based quantum computing (MBQC) and building practical quantum networks. One conventional way to generate multipartite entangled state is combining optical parametric processes\cite{073601} and BSs\cite{167,96}, but the partite number is limited by the complex experimental setup and technical difficulty. Another effective way is using frequency/time multiplexing\cite{124205,7535}, by which multipartite entanglement of ultra-large scale can be achieved with a more compact setup. 
However, hybrid quantum networks depend on the generation and distribution of entanglement among different physical systems (with different frequencies), such as atoms\cite{1098,093601}, ions\cite{153,281}, quantum dots\cite{426}, superconducting circuits\cite{20171,589}, etc., which is a big challenge for the above-mentioned methods. Therefore, the optomechanical system is considered as a promising candidate for connecting different components in hybrid quantum systems. Thus, various entangled states have been generated from cavity optomechanical systems theoretically\cite{042330,17237,042342,054061,29581,063602,0485,559,053515,15032,063801,1,485,052303,033842,10306,042320} and experimentally\cite{1244563,651,478,473}, including entangled states between light (microwave) modes\cite{17237,042330,054061,042342}, light (microwave) and mechanical modes\cite{651,1244563,29581,063602,0485,559}, and mechanical modes\cite{478,473,053515,15032}. In particular, entanglement between light and/or microwave modes generated from an optomechanical system is of great importance, because it is valuable for building hybrid quantum networks. In the past ten years, entanglement between two light modes\cite{17237,063801,1,485,052303,054061,042342}, such as two spatial modes\cite{063801,1,485}, or two longitudinal modes in a single cavity optomechanical system\cite{052303}, two cavity modes in a double-cavity optomechanical system\cite{17237}, were investigated and discussed.
A scheme of entangling optical and microwave cavity modes by means of a nanomechanical resonator was also proposed, which indicates that the mechanical resonator can mediate the robust entanglement between the optical and microwave cavity modes\cite{042342}. The generation of microwave-optical entanglement from a piezo-optomechanical system and microwave-microwave entanglement by entanglement swapping with the generated entangled microwave-optical sources was studied recently\cite{054061}. 
In addition, generating multipartite entanglement through interaction with the mechanical mode, including three optical cavity modes, two optical and one microwave modes, one optical and two microwave modes, were further proposed and discussed\cite{033842,10306,042320}. However, in the above mentioned works, the partite numbers of the entangled state among optical or/and microwave modes were no more than three, which is not enough for connecting many different “notes” composed of different physical systems, in a hybrid quantum network. In this paper, we discuss the double-longitudinal-mode cavity optomechanical system in detail and propose two schemes for generating a multipartite continuous variable (CV) entangled state by coupling some certain output modes of N such systems.

\section{Double-longitudinal-mode Cavity Optomechanical System}
\begin{figure}[htbp]
\centering\includegraphics[width=10cm]{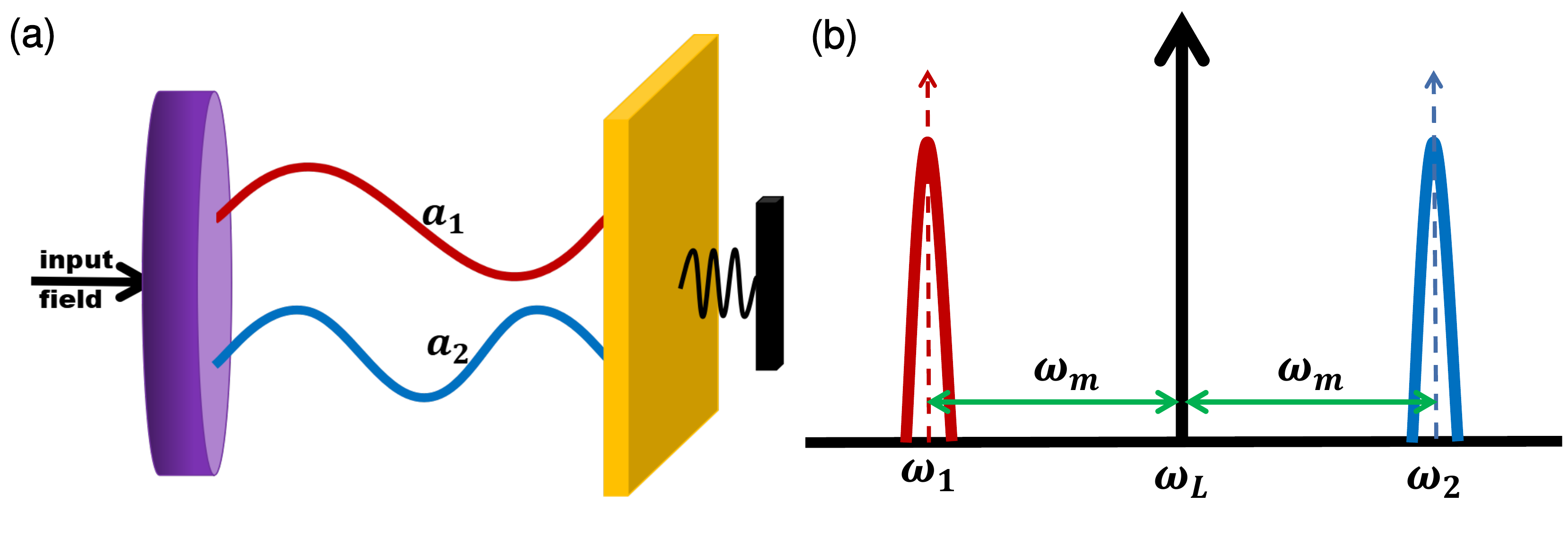}
\caption{The considered double-longitudinal-mode cavity optomechanical system.\\     (a) Schematic diagram. (b) frequency ralation.}
\end{figure}
The driven double-longitudinal-mode cavity optomechanical system is shown in Fig.1(a). The cavity includes a fixed mirror and a movable mirror (with frequency $\omega_m$ and decay $\gamma_m$) and is driven by an input laser (with frequency $\omega_L$). Considering the condition that the mechanical frequency $\omega_m$ is equal to half of the free spectral range (FSR), the corresponding Stokes and anti-Stokes side-bands of the input driven laser can resonate with two longitudinal modes of the cavity, and the frequency relation is shown in Fig.1(b). The Hamiltonian of this system can be described as:

{\begin{equation}
    \begin{aligned}
  \hat{H}  &= \hbar\omega_1 \hat{a}^{\dagger}_1 \hat{a}_1+\hbar\omega_2 \hat{a}^{\dagger}_2 \hat{a}_2+\hbar\frac{\omega_m}2(\hat{q}^2+\hat{p}^2) - \hbar g_0(\hat{a}_1+\hat{a}_2)^{\dagger}(\hat{a}_1+\hat{a}_2)\hat{q} \quad \\
  &\quad + i\hbar\eta_1(\hat{a}^{\dagger}_1e^{-i\omega_Lt}- \hat{a}_1e^{i\omega_Lt})+i\hbar\eta_2(\hat{a}^{\dagger}_2 e^{-i\omega_{L} t}- \hat{a}_2 e^{i\omega_{L} t}),
    \end{aligned}
\end{equation} }
where $\hat{a}^{\dagger}_j$ ($\hat{a}_j$) ($j=1, 2$) is the creation (annihilation) operator of the cavity mode $j$ (with frequency $\omega_j$ and decay rate $\kappa_j$ ). $\hat{p}$ and $\hat{q}$ are the dimensionless momentum and position operators of the mechanical mode. The optomechanical coupling coefficient $g_0=\sqrt{\hbar\omega_1\omega_2/m\omega_m}/L$, where $L$ is the cavity length and $m$ is the effective mass of the mechanical oscillator. $\eta_j=\sqrt{2P\kappa_j/\hbar\omega_j}$ denotes the coupling strength between driving field and cavity fields, where $P$ is the input power. The first three terms of the Eq.(1) are the free Hamiltonians of the cavity mode 1, and 2 and the movable mirror, respectively. The fourth term represents the optomechanical coupling, and the last two terms are the driving terms of the cavity modes.

From the Hamiltonian described by Eq.(1), the linearized quantum Langevin equations of the quadrature fluctuations ($\delta q$, $\delta p$, $\delta X_{1}$, $\delta Y_{1}$, $\delta X_{2}$, $\delta Y_{2}$ ) can be obtained as $\dot{u}=\mathcal{A} u + n(t)$ (details are shown in Appendix 5.1), where $\delta X_{j}=(\delta a_{j}+\delta a_{j}^{\dagger})/\sqrt{2}$, and $\delta Y_{j}=i(\delta a_{j}^{\dagger}-\delta a_{j})/\sqrt{2}, j=1, 2$. Then the corresponding covariance matrix $CM$ can be obtained by solving the Lyapunov equation, and the optical-optical and optomechanical entanglements were studied carefully in Ref\cite{052303}, with the criterion of logarithmic negativity. $\Delta_{j}=\omega_{j}-\omega_{L} (j=1, 2)$ is the frequency detuning of the cavity field $i$ with respect to the input laser field in the frame rotating at the input field frequency $\omega_L$. However, the steady state solutions used in Ref\cite{052303} are approximate and some special characteristics were absent. Therefore, we check the model again and further investigate the parameter dependence (including the detuning $\Delta_{2}$, cavity decay $\kappa$, mechanical decay $\gamma_{m}$ and temperature $T$) of the entanglements, based on the general steady state solutions of the quantum Langevin equations. 
\begin{figure}[htbp]
\centering\includegraphics[width=7cm]{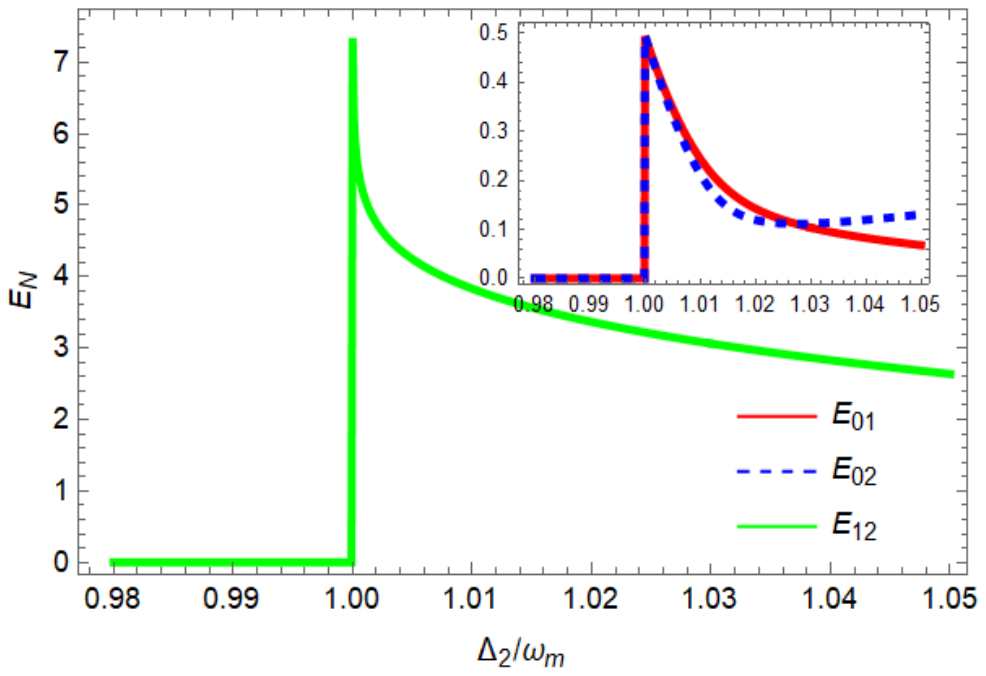}
\caption{The optical-optical entanglement $E_{12}$, the optomechanical entanglements $E_{01}$ and $E_{02}$, versus $\Delta_{2}$. The relative parameters are: $L=0.01m$, $T= 0.01 K$, $\lambda=1.33\mu m$, $\gamma_{m}=0.1MHz$, $\kappa_{1}=\kappa_{2}=\kappa=1MHz$, $m=5\times10^{-9} kg$, $P=20mW$, $\Delta_{1}=\Delta_{2}-2\omega_{m}$. }
\end{figure}
The optical-optical entanglement $E_{12}$, the optomechanical entanglements $E_{01}$ and $E_{02}$, versus detuning $\Delta_2$,  are shown in Fig.2. As expected, all optical-optical and optomechanical entanglements reach their maximum at $\Delta_{2}=\omega_{m}$ ($\Delta_{1}=-\omega_{m}$), due to the perfect collaboration of the two-mode squeezing (entanglement generation) and the beam-splitter interaction (state transfer) mechanisms. With the increase of detuning $\Delta_{2}$, Stokes and anti-Stokes scattering processes become unbalanced and these two mechanisms are weaker and the optical-optical entanglement $E_{12}$ decreases. Note that the two optomechanical entanglements are not identical, which is different from that indicated in Ref\cite{052303}. In particular, entanglement between movable mirror and cavity mode 2 is not monotonous, it experiences a decreasing period with smaller entanglement and then an increasing period with larger entanglement, compared with entanglement between the mirror and the cavity mode 1. This is understandable: for $E_{01}$, the entanglement generation mechanism changes from strong to weak, while for $E_{02}$, the cooling mechanism becomes to dominate after a period \cite{1391}, which can react the entanglement to be better, during the increase of detuning $\Delta_2$. It is also noticed that all entanglements only exist in the zone $\Delta_{2}\geq\omega_{m}$, due to the negative effective mechanical damping rate $\Gamma_{eff}$ at $\Delta_{2}<\omega_{m}$, which leads to amplification of thermal fluctuations and finally to an instability and the prediction of entanglement is breakdown\cite{1391}, under the parameters of Fig.2. The definition of the effective mechanical damping rate is $\Gamma_{eff} =\Gamma_{opt} +\Gamma_{m}$, where $\Gamma_{opt}$ is the optomechanical damping rate and $\Gamma_{m}$ is the mechanical damping rate.  $\Gamma_{eff}<0$ at $\Delta_{2}<\omega_{m}$, because $\Gamma_{opt}$ is negative and $\Gamma_{m} \ll \lvert \Gamma_{opt} \rvert$ under the conditions here. 
\begin{figure}[htbp]
\centering\includegraphics[width=13cm]{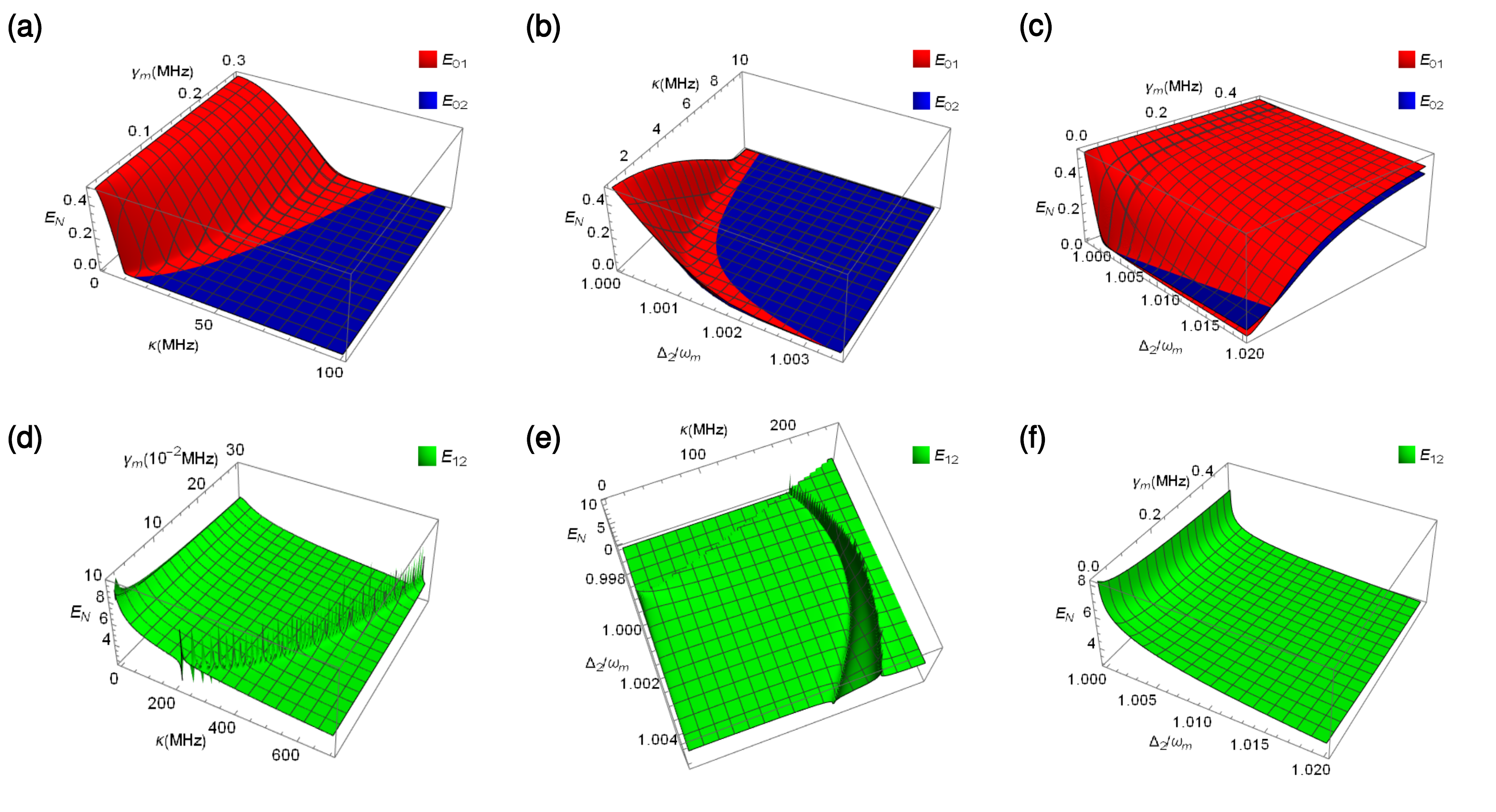}
\caption{The optomechanical entanglements $E_{01}$ and $E_{02}$ ((a),(b),(c)), and the optical-optical entanglement $E_{12}$ ((d),(e),(f)), vary with detuning $\Delta_2$, cavity decay $\kappa$ and mechanical decay $\gamma_m$. In some regions of subgraph (a), (b) and (c), only one color is shown and the visible color stands for stronger entanglement. $\gamma_{m}=0.01MHz$. Other parameters are the same as that in Fig.2.}
\end{figure}
The optomechanical entanglements $E_{01}$ and $E_{02}$ versus cavity delay $\kappa$, mechanical delay $\gamma_m$, and detuning $\Delta_{2}$, are shown in Fig.3(a)-(c). It is obvious that optomechanical entanglements decrease with the increase of cavity decay $\kappa$, while increasing $\gamma_m$ can slow down this decrease of entanglement, as is shown in Fig.3(a). In Fig.3(b), optomechanical entanglements decrease with the increase of detuning $\Delta_{2}$ and increasing $\kappa$ can fasten this decrease of entanglement. In Fig.3(c), increasing $\gamma_m$ can keep the entanglement robust when $\Delta_{2}=\omega_m$ (maximum entanglement condition), and can promote the entanglement to bigger value when $\Delta_{2}>\omega_m$. These results means that bigger detuning or cavity decay will destroy the entanglements, but bigger mechanical decay $\gamma_m$ may help to increase the entanglements and there exists an optimized mechanical decay, as is further shown in Fig.4. The optical-optical entanglement $E_{12}$ versus cavity delay $\kappa$, mechanical delay $\gamma_m$, and detuning $\Delta_{2}$, are shown in Fig.3(d)-(f). In Fig.3(d), $E_{12}$ decrease when $\kappa$ and $\gamma_m$ increase. It is interesting that there is a dispersion-like shape (corresponding to the two-phonon resonance) at a certain point of $\kappa$, because the Lorentz line shape of two cavity modes cannot be ignored if $\kappa$ is big enough. Increasing $\gamma_m$ makes the position of the certain point of $\kappa$ change to a bigger value, because faster decaying phonon number decreases the probability of the two-phonon resonance. Fig.3(e) shows that bigger detuning $\Delta_2$ makes the dispersion-like region wider, because frequency difference between cavity mode 2 and pump mode increases and two cavity modes are no longer symmetric to the pump mode any more. Note that nonzero entanglement appears at $\Delta_2<\omega_m$, due to stronger cavity decay $\kappa$ (resulting in smaller $\lvert \Gamma_{opt} \rvert$), which is different from that in Fig.2. Fig.3(d)-(f) prove that detuning, mechanical or cavity decays, are all negative factors for optical-optical entanglement. However, mechanical decay can be a positive factor for optomechanical entanglements, which is counter-intuitive. Therefore, larger ranges of $\gamma_m$ are considered and shown in Fig.4(a) and Fig.4(b), with different detunings. The positive role of mechanical decay $\gamma_m$ is limited, and optimized values (changes with detuning) exist for the strongest entanglements.
\begin{figure}[htbp]
\centering\includegraphics[width=10cm]{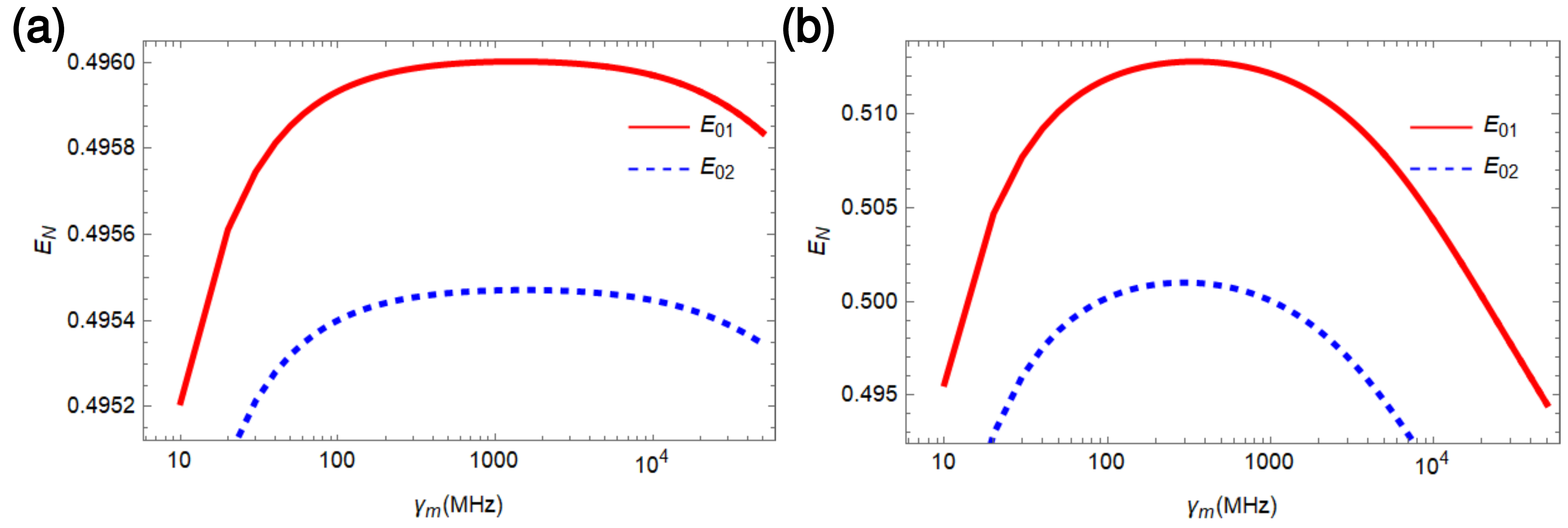}
\caption{Optomechanical entanglements $E_{01}$ and $E_{02}$ versus mechanical decay $\gamma_m$. (a) $\Delta_{2}=\omega_m$. (b)$\Delta_{2}=1.005\omega_m$. $\kappa=10MHz$. Other parameters are the same as that in Fig.2.}
\end{figure}
\begin{figure}[htbp]
\centering\includegraphics[width=10cm]{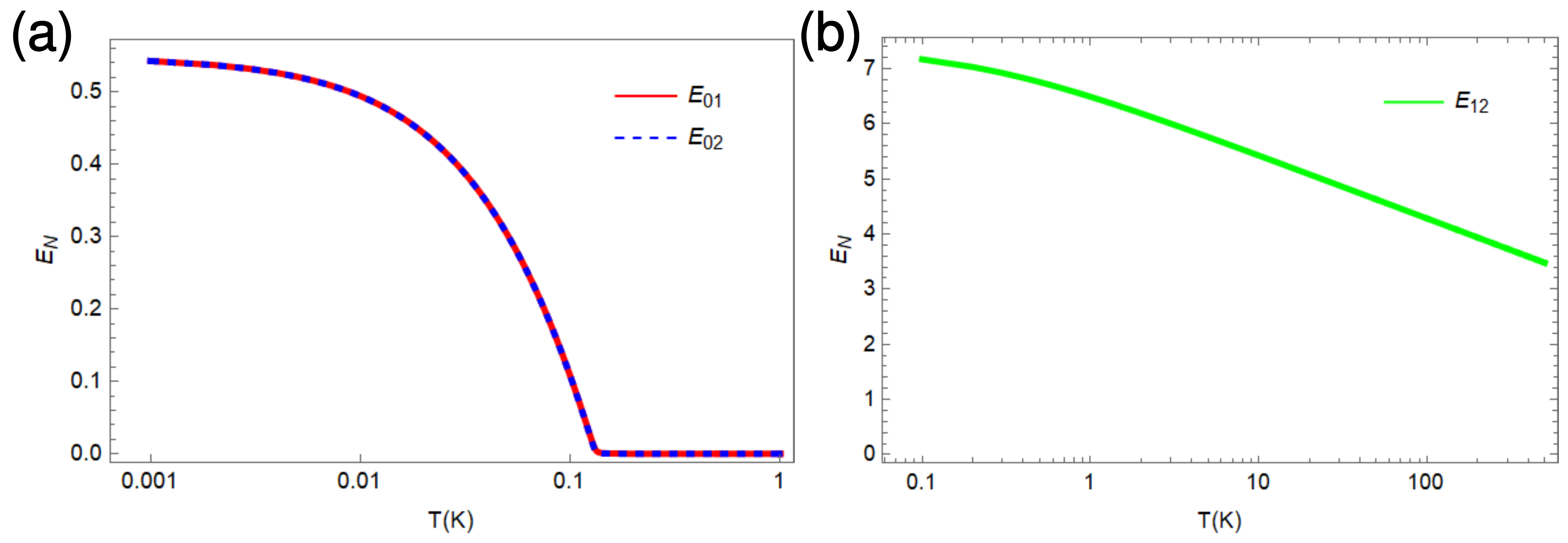}
\caption{Entanglements versus temperature $T$, with $\Delta_{2} = \omega_{m}$ and $\gamma_{m}=0.1MHz$. (a) optomechanical entanglements. (b) optical-optical entanglement. Other parameters are the same as that in Fig.2. }
\end{figure}

In addition, other than the optomechanical entanglements, the optical-optical entanglement is robust enough even at room or higher temperatures, as is shown in Fig.5, which is the same as the corresponding conclusion in Ref\cite{052303}.

\section{Multipartite optical-optical entanglement generation}
To realize multipartite entanglement is usually to increase the dimension of entanglement, because the quantity of entanglement is finite in each dimension. Based on the robust bipartite optical-optical entanglement, one can further expand the partite number (which is very important to the ability of quantum computing) generally in two ways: coupling them with BS(s)\cite{96}, and using double pumps with orthogonal polarization\cite{0485,559}. In the following sections, these two ways are discussed separately to generate multipartite optical-optical entanglement. To make it clearer and more understandable, schemes for generating quadrapartite entanglements are considered first.

\subsection{Scheme 1: Coupling with BS(s)}
\begin{figure}[htbp]
\centering\includegraphics[width=10cm]{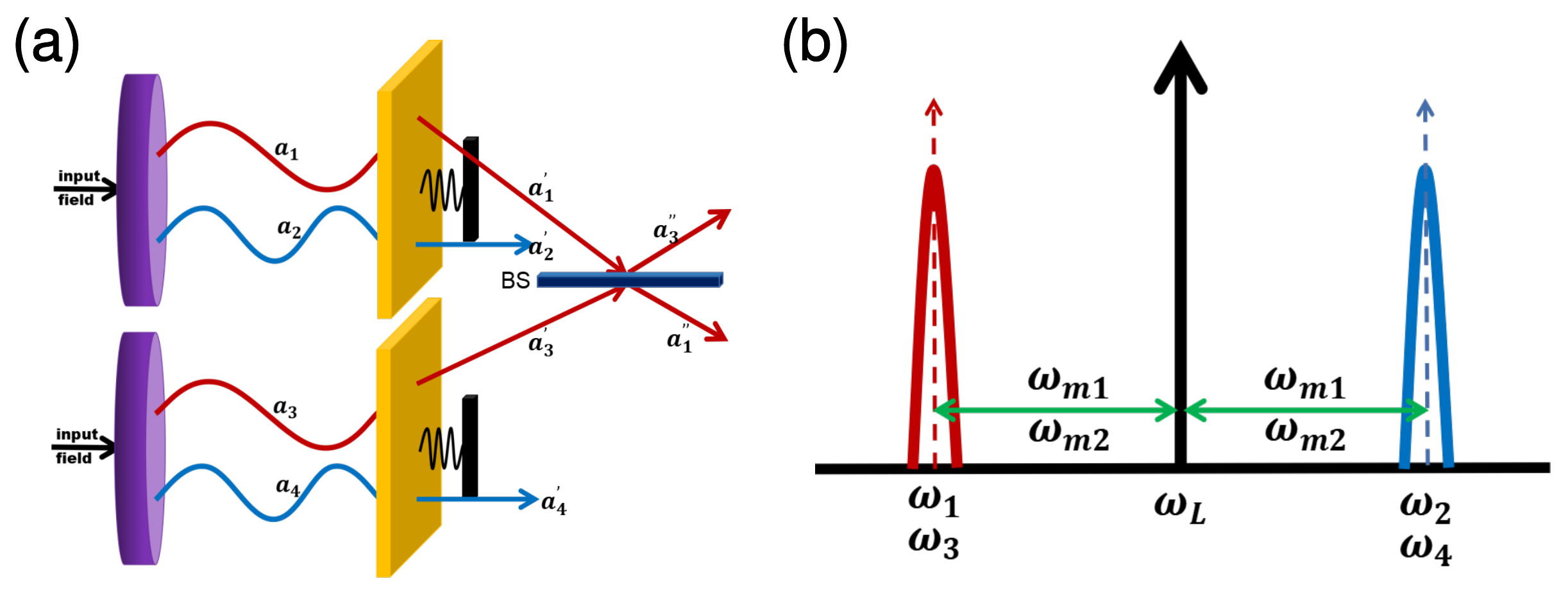}
\caption{The scheme of generating quadrapartite entanglement using a BS.\\     (a) Schematic diagram. (b) frequency ralation.}
\end{figure}
The scheme of generating quadrapartite entanglement by using a BS, is shown in Fig.6(a). Output beams $a^{\prime}_1$ and $a^{\prime}_3$ (same frequency) from two identical double-longitudinal-mode cavity optomechanical systems, are coupled through a BS, and the corresponding frequency relations are shown in Fig.6(b). Here $a^{\prime}_1$ and $a^{\prime}_2$ are the two output modes of $a_1$ and $a_2$, and $a^{\prime}_3$ and $a^{\prime}_4$ represent the output modes of the added optomechanical cavity, where $\omega_1=\omega_3$, $\omega_2=\omega_4$, and $\omega_{m1}=\omega_{m2}=\omega_m$. In this case, the Hamiltonians of the two cavity optomechanical systems can be described separately:

{\begin{equation}\begin{aligned}
 \hat{H}_{1}&=\sum_{j=1,2} \hbar\omega_{j} \hat{a}^{\dagger}_{j} \hat{a}_{j}+\hbar\frac{\omega_{m1}}2(\hat{q}^2_{1}+\hat{p}^2_{1})-\hbar g_{0}(\hat{a}_{1}+\hat{a}_{2})^{\dagger}(\hat{a}_{1}+\hat{a}_{2})\hat{q}_{1} \quad \\ &\quad +\sum_{j=1, 2} i\hbar\eta_{j}(\hat{a}^{\dagger}_{j}e^{-i\omega_{L} t}- \hat{a}_{j}e^{i\omega_{L} t}),
\end{aligned}\end{equation}}
{\begin{equation}
\begin{aligned}
\hat{H}_{2}&=\sum_{j=3, 4} \hbar\omega_{i} \hat{a}_{j}^{\dagger} \hat{a}_{j}+\hbar\frac{\omega_{m2}}2(\hat{q}^2_{2}+\hat{p}^2_{2})-\hbar g_{0}(\hat{a}_{3}+\hat{a}_{4})^{\dagger}(\hat{a}_{3}+\hat{a}_{4})\hat{q}_{2} \quad \\ &\quad +\sum_{j=3, 4} i\hbar\eta_{j}(\hat{a}^{\dagger}_{j}e^{-i\omega_{L} t}- \hat{a}_{j}e^{i\omega_{L} t}).
\end{aligned}
\end{equation}}

Here $g_0=\sqrt{\hbar \omega_{1} \omega_{2}/ m \omega_{m1}} / L =\sqrt{\hbar \omega_{3} \omega_{4}/ m \omega_{m2}} / L $, and $\eta_{j}=\sqrt{2P\kappa_{j}/\hbar\omega_{j}}$. Other parameters are defined the same as the above. The third terms represent the optomechanical couplings, and the last terms are the driving terms of the cavity modes. As is well known, the BS transformation can be expressed by matrix $U _{BS}=\left(\begin{matrix}
   cos\theta e^{i\phi} & sin\theta e^{i(\phi+\pi)} \\
   sin\theta  & cos\theta \\
\end{matrix} \right)$, where $\phi$ and $cos\theta$ represent the relative phase  and the transmission coefficient of the BS. 
\begin{figure}[htbp]
\centering\includegraphics[width=13cm]{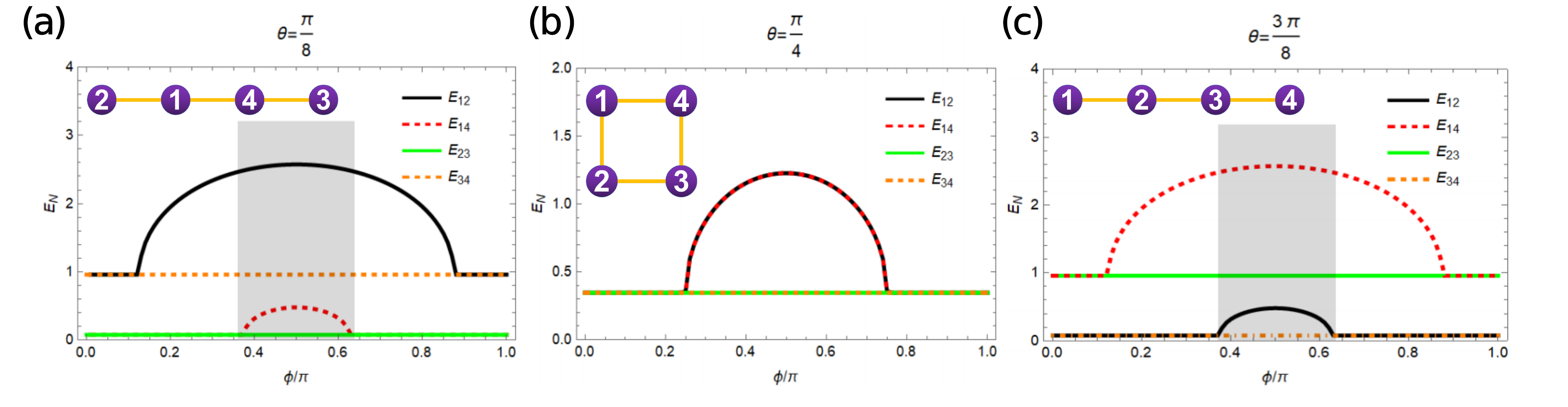}
\caption{The quadrapartite entanglement versus BS parameters $\phi$ with different $\theta$. $L=0.01m$, $T= 0.01 K$, $\omega_{m}=\frac{\pi c}{2L}$, $\lambda=1.33\mu m$, $\kappa=100\gamma_{m}=1MHz$, $m=5\times10^{-9} kg$, $P=20mW$, $\Delta_{1}=-\Delta_{2}=\Delta_{3}=-\Delta_{4}=-\omega_{m}$.}
\end{figure}
Therefore, the generated quadrapartite entanglment versus BS parameters ($\phi$ and $\theta$) can be obtained (details can be found in Appendix 5.2), as is shown in Fig.7. It is interesting that one can obtain different quadrapartite entanglement structures by changing the BS parameters. When $\theta=\pi/8$ (85:15) and $\theta=3\pi/8$ (15:85), quadrapartite entanglements with linear structure can be obtained at $3\pi/8<\phi<5\pi/8$ (described by the gray shadow areas), as are shown in Fig.7(a) and Fig.7(c), and note that the two linear structures are not the same. If $\theta=\pi/4$ (50:50), a square structure entanglement can be generated, regardless of the relative phase $\phi$ takes, as is shown in Fig.7(b). Therefore, this scheme may help to realize controllable multipartite entanglement, whose entanglement structure can be adjusted by changing the BS parameters.

\begin{figure}[htbp]
\centering\includegraphics[width=13cm]{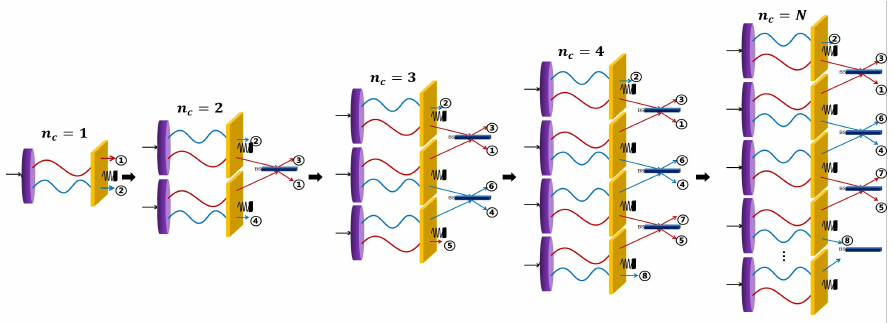}
\caption{Generation of 2N-partite entanglement. $n_c$ is the number of cavities used in the scheme.}
\end{figure}

\begin{figure}[htbp]
\centering\includegraphics[width=13cm]{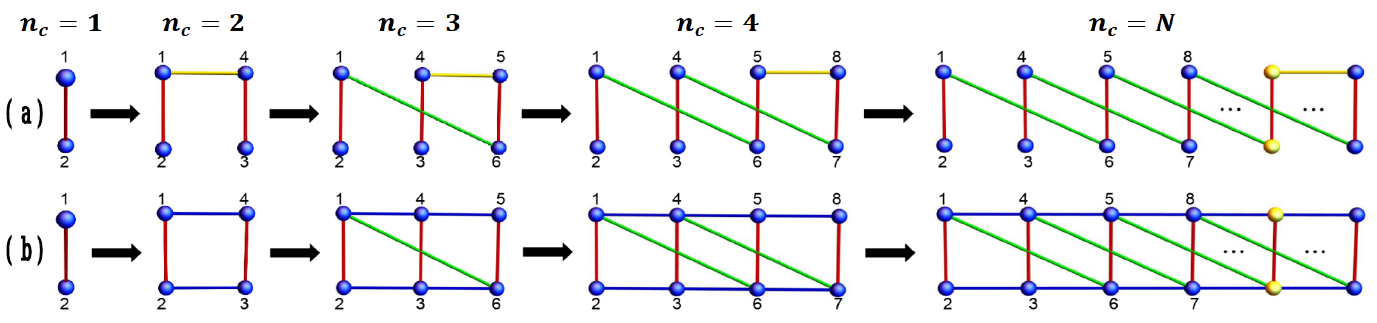}
\caption{Generation of linear (a) and “double ladder” (b) entanglement structure. (a) $\theta=\pi/8$ and $\phi=\pi/2$. (b) $\theta=\pi/4$ and $\phi=\pi/2$. Other parameters are the same as that in Fig.7. Note that the blue balls stand for those optical modes, and each column corresponds to two modes from the same cavity. The column formed by two yellow balls stand for the middle part that are not shown separately.}
\end{figure}
Moreover, 2N-partite entanglement generation scheme based on the above scheme is shown in Fig.8. By comparing many identical double-longitudinal-mode cavity optomechanical systems and coupling output beams with same frequencies through several BSs, a 2N-partite entangled state can be generated, and the corresponding entanglement structures with different BS parameters are shown in Fig.9. As is shown in Fig.9(b), the entanglement structure changes from bipartite (one cavity) to quadrapartite (two cavities), hexapartite (three cavities), octapartite (four cavities), ..., and 2N-partite (N cavities), by using 0, 1, 2, 3, ..., N-1 50:50 BSs, respectively. Here red bars describe the original entanglements from double-longitudinal-mode cavity optomechanical systems, and blue and green bars mean the generated entanglements between modes of adjacent and next-adjacent cavities, respectively, after using BS(s). The generated 2N-partite entanglement has a special  “double ladder” structure, including ladders formed by red-blue and blue-green bars. In Fig.9(a), different BS parameter ($\theta=\pi/8$) is considered, and a linear entanglement structure can be obtained. Note that this linear structure can be viewed as removing those blue bars in the “double ladder” structure shown in Fig.9(b), and the yellow bar describes the entanglement generated from the last added BS. Therefore, control of the entanglement structure can be realized by changing the BS parameters, for further applications of quantum computing.

\subsection{Scheme 2: Using Double Pumps with Orthogonal Polarization}

\begin{figure}[htbp]
\centering\includegraphics[width=10cm]{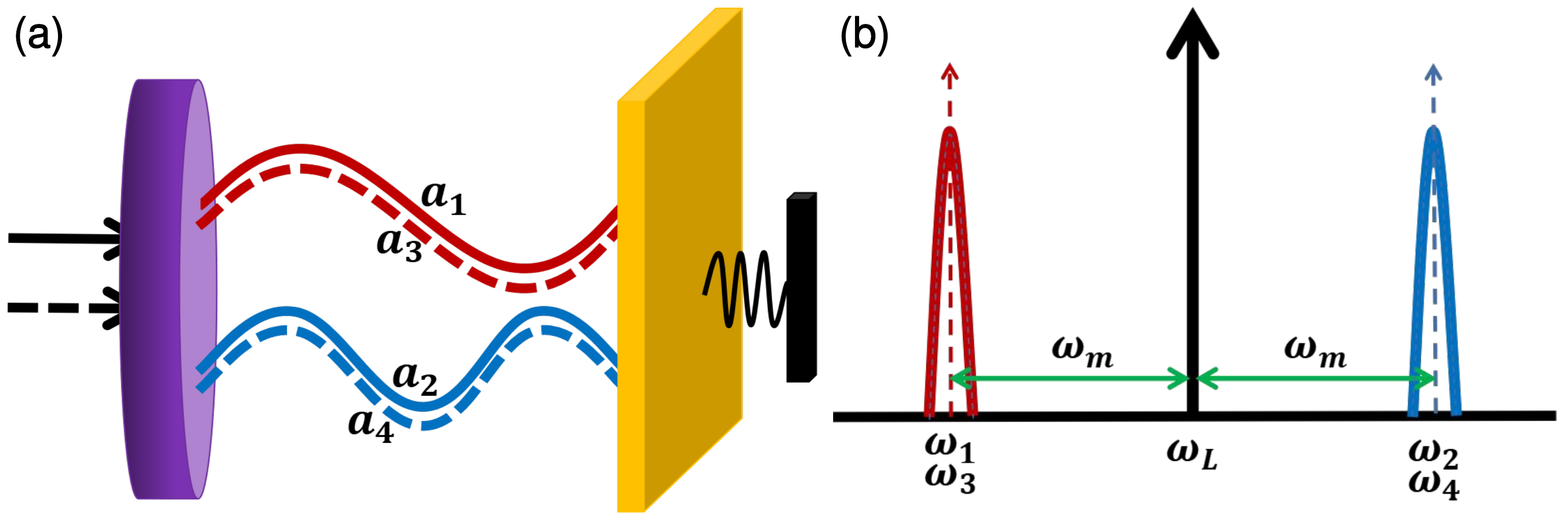}
\caption{The scheme of generating quadrapartite entanglement by using double pumps with orthogonal polarizations.\\     (a) Schematic diagram. (b) frequency ralation.}
\end{figure}

The other scheme of generating quadrapartite entanglement by using double pumps with orthogonal polarizations is shown in Fig.10(a), and Fig.10(b) indicates the corresponding frequency relation. This time two sets of longitudinal modes with orthogonal polarizations both resonate in the cavity. The intracavity modes $a_1$ ($a_2$) and $a_3$ ($a_4$) have same frequencies, $\omega_1=\omega_3$ ($\omega_2=\omega_4$), but their polarizations are different ($a_1$ ($a_2$): H, $a_3$ ($a_4$): V). Then, the Hamiltonian of this system can be described as\cite{113601,120602,023812}:
{\begin{equation}
\begin{aligned}
\hat{H}= & \sum_{j=1}^{4} \hbar\omega_{j} \hat{a}^{\dagger}_{j} \hat{a}_{j} + \hbar\frac{\omega_{m}}2(\hat{q}^2+\hat{p}^2)-\hbar g_0 (\hat{a}_{1}+\hat{a}_{2})^{\dagger} (\hat{a}_{1}+\hat{a}_{2}) \hat{q} - \hbar g_0(\hat{a}_{3}+\hat{a}_{4})^{\dagger}(\hat{a}_{3}+\hat{a}_{4})\hat{q}\\ & +\sum_{j=1}^4 i\hbar\eta_{j} (\hat{a}^{\dagger}_{j} e^{-i\omega_{L} t}- \hat{a}_{j} e^{i\omega_{L} t}).
\end{aligned}
\end{equation}}

Here $g_0=\sqrt{\hbar \omega_{1} \omega_{2}/ m \omega_{m}} / L =\sqrt{\hbar \omega_{3} \omega_{4}/ m \omega_{m}} / L $, and $\eta_{j}=\sqrt{2P\kappa_{j}/\hbar\omega_{j}}$. Other parameters are defined the same as the above. 

\begin{figure}[htbp]
\centering\includegraphics[width=10cm]{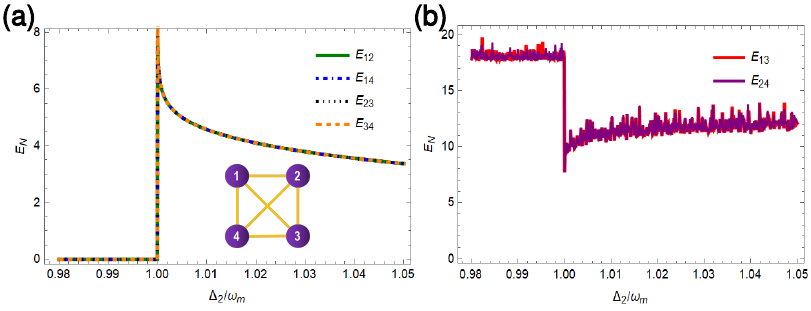}
\caption{The optical-optical entanglements with different frequency ($E_{12}$, $E_{34}$, $E_{14}$ and $E_{23}$) and with same frequency ($E_{13}$ and $E_{24}$) versus detuning $\Delta_{2}$. The relative parameters are the same as that in Fig.2.}
\end{figure}

Based on scheme 2, entanglements between cavity modes can be obtained (details can be found in Appendix 5.3), as is shown in Fig.11. Fig.11(a) and Fig.11(b) show that entanglement can be generated between each two of the four optical modes, i.e., a quadrapartite GHZ entangled state can be obtained. Specifically, entanglements between those modes with different frequencies $E_{12}$, $E_{34}$, $E_{14}$ and $E_{23}$ are the same and have the same value as those in single double-londitudinal-mode cavity optomechanical system, due to the same entanglement generation mechanism. However, entanglements between modes with the same frequency ($E_{13}$ and $E_{24}$) have totally opposite trend and are not limited by the effective mechanical damping rate. This may result from different mechanisms: connecting modes with same frequency by mechanical mode is different from the Stokes and anti-Stokes scattering processes that connecting modes with different frequencies. It seems that the mechanical mode (the movable mirror) is not so sensitive to polarization, but has a strong dependence on frequency. Note that the entanglements $E_{13}$ and $E_{24}$ always fluctuate with the detuning $\Delta_2$, which is similar to that in Ref\cite{17237}. The fluctuating amplitude is related to the cavity and mechanical decays, and a larger decay will lead to more violent fluctuations.

\begin{figure}[htbp]
\centering\includegraphics[width=13cm]{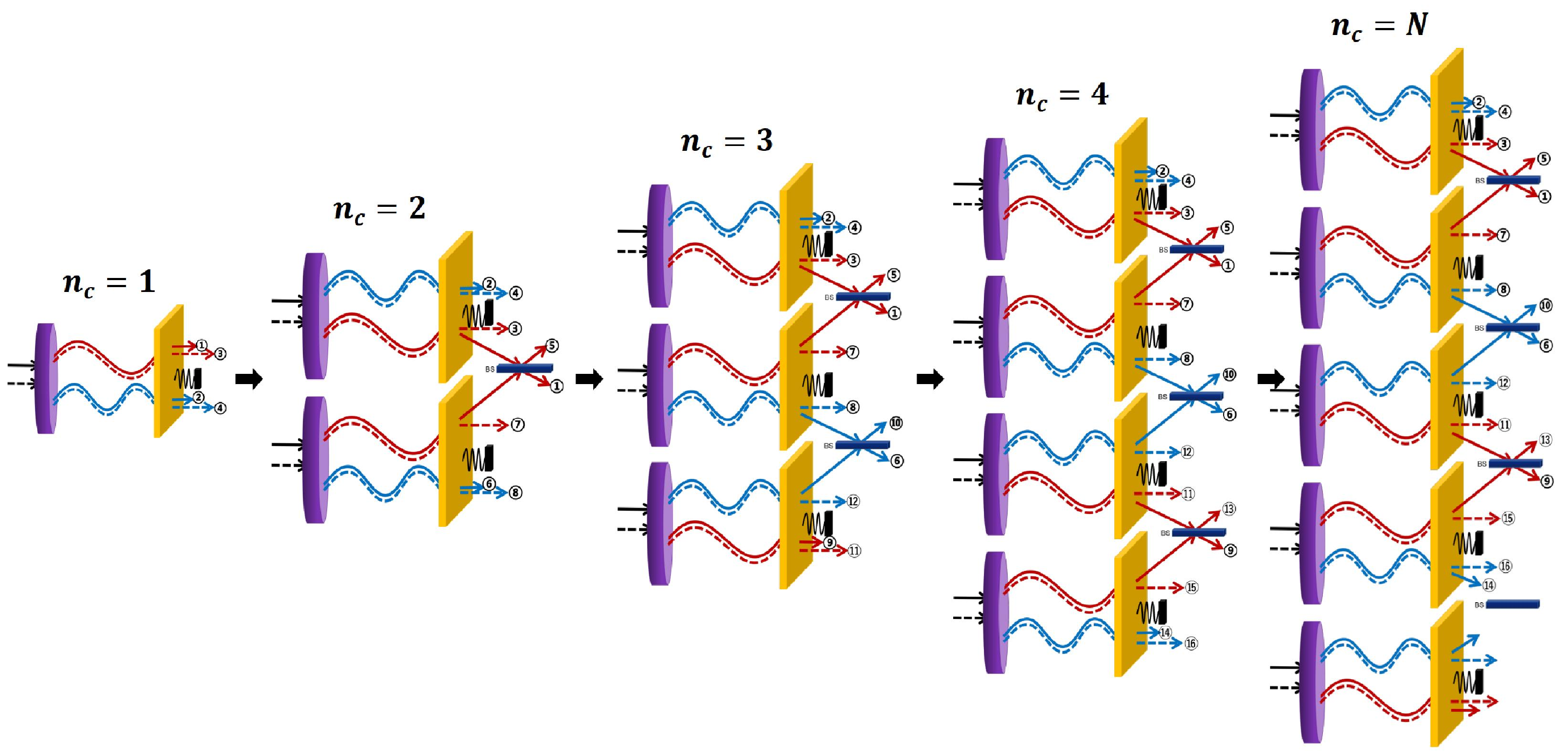}
\caption{Generation of 4N-partite entanglement. $n_c$ is the number of cavities used in the scheme.}
\end{figure}

\begin{figure}[htbp]
\centering\includegraphics[width=13cm]{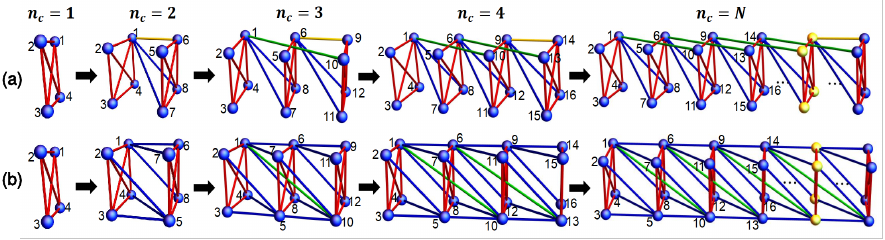}
\caption{(a) $\theta=\pi/8$ and $\phi=\pi/2$. (b) $\theta=\pi/4$ and $\phi=\pi/2$. Other parameters are shown as Fig.11. Note that the blue balls stand for those optical modes, and each slice corresponds to four modes from the same cavity. The slice formed by four yellow balls stand for the middle part that are not shown separately.}
\end{figure}
Similarly, by coupling certain output beams from several such kind optomechanical cavities pumped with orthogonal polarizations, a 4N-partite entangled state can be obtained, as is shown in Fig.12 and Fig.13. Here red bars still describe the original entanglements from orthogonal-polarization-pumped optomechanical systems, and blue and green bars still mean the generated entanglements between modes of adjacent and next-adjacent cavities, respectively, similar to that in Fig.9. Here the generated 4N-partite entanglement by using 50:50 BSs, as is shown in Fig.13(b), contains 6 sets of ladder entanglement structures including 5 ladders formed by red-blue bars and 1 ladder formed by blue-green bars. In Fig.13(a), different BS parameter ($\theta=\pi/8$) is considered, and a special entanglement structure can be obtained, which can be viewed as removing some blue bars in the structure shown in Fig.13(b), and the yellow bar still describes the entanglement generated from the last added BS. Thus, one can control this kind of three-dimensional entanglement structure by changing BS parameters, which provides rich possibilities for building some special-structured 3D multipartite entangled states towards quantum computing and quantum networks.

\section{Conclusion}
In this paper, we investigate the parameter dependence of optomechanical and optical-optical entanglements generated from double-longitudinal-mode cavity optomechanical system and propose two multipartite entanglement generation schemes based on such systems. By coupling N cavities with N-1 BSs, 2N or 4N-partite entangled states can be obtained, which contain certain ladder or linear structures and may be promising choices to be applied in quantum computing and quantum networks in the future. Our schemes of increasing entanglement-partite-number have high scalability, and the obtained entanglement can be further enhanced by cascading, feedback and nonlinear gain medium. Considering this optomechanical coupling together with multiplexing of new degrees of freedom, such as frequency, orbital angular momentumn, etc., it is also possible to build large-scale entangled states, which is very important for quantum computing ability. Our work provides a new thought of generating multipartite entangled states applied to connecting different physical systems with different frequencies in practical quantum networks.\\

\begin{backmatter}
\bmsection{Funding}
National Key Research and Development Program of China (2021YFA1402002, 2021YFC2201802); Natural Science Foundation of China (NSFC) (11874249, 11874248, 11974225, 12074233).
\bmsection{Disclosures}
The authors declare no conflicts of interest.
\bmsection{Data availability} Data underlying the results presented in this paper are not publicly available at this time but may be obtained from the authors upon reasonable request.
\end{backmatter}

\section{Appendix}
Here we provide the details on how we obtain the steady state entanglement. The entanglement is calculated based on the covariance matrix of the cavity modes and mechanical mode. The covariance matrix can be achieved by solving the quantum Langevin equations, which can be rewritten in the following form:
\begin{equation}
\dot{u}(t)=\mathcal{A} u(t)+n(t)
\end{equation}
where u(t) is the vector of quadrature fluctuation operators of cavity modes and mechanical mode. $\mathcal{A}$ is the drift matrix, and n(t) is the vector of noise quadrature operators associated with the noise terms. 

The steady-state covariance matrix $V(t\rightarrow\infty)$ of the system quadratures, with its entries defined as $V_{ij} = \frac{1}{2} \langle{\{u_i(t),u_j(t)\}}\rangle$, which can be obtained by solving the Lyapunov equation:
\begin{equation}
\mathcal{A}V+V\mathcal{A}^{T}=-\mathcal{D}
\end{equation}
where $\mathcal{D}$ is the diffusion matrix, with its entries defined as
$\frac{1}{2} \langle{n_i(t) n_j(t^{\prime}) +n_j(t^{\prime}) n_i(t)}\rangle=\mathcal{D}_{ij}\delta(t-t^{\prime})$. $a_{in}$ and $\xi$ are the optical noise operators of the cavity mode and the mechanical mode, respectively, which are zero mean and characterized by the following correlation function: 

{\begin{equation}\begin{aligned}
\left \langle a_{in}(t)a_{in}^{\dagger}(t^{\prime}) \right \rangle&=\delta(t-t^{\prime})\\\left \langle \xi(t)\xi(t^{\prime})+\xi(t^{\prime}) \xi(t)\right \rangle &\simeq\gamma_{mi}(2{\overline{n}}+1)\delta(t-t^{\prime})                         
\end{aligned}\end{equation}}
where a Markovian approximation has been made, and ${\overline{n}}=\left[exp[(\hbar \omega_{m}/k_{B}T)]-1\right]^{-1}$ is the equilibrium mean phonon.

Once the covariance matrix $V$ is obtained, the entanglement can then be quantified by means of logarithmic negativity\cite{032314}:
\begin{equation}
E_N=max[0, -ln2\nu_{-}]
\end{equation}
where $\nu_{-} = min \thinspace eig\lvert i\Omega_2V_m\rvert$ ($\Omega_2= \oplus i \sigma_y $ is the so-called symplectic matrix and $\sigma_y$ is the y Pauli matrix) is the minimum symplectic eigenvalue of the covariance matrix $V_m= PVP$, with $V_m$ being the $4 \times 4$ covariance matrix associated, and $P = diag\begin{pmatrix}
1, 1, 1, -1
\end{pmatrix}$ the matrix that inverts the sign of momentum, $p_2\rightarrow -p_2$, realizing partial transposition at the level of covariance matrices.

\subsection{Single Double-longitudinal-mode Cavity Optomechanical System}

From the Hamiltonian of a single double-longitudinal-mode cavity optomechanical system, which is described in Eq.(1), the corresponding linearized $QLEs$ of quadrature fluctuations ($\delta q$, $\delta p$, $\delta X_{1}$, $\delta Y_{1}$, $\delta X_{2}$, $\delta Y_{2}$) can be obtained as:
{\begin{equation}
\begin{pmatrix}
\delta \dot{q} \\ \delta \dot{p} \\ \delta \dot{X}_{1} \\ \delta \dot{Y}_{1} \\ \delta \dot{X}_{2} \\ \delta \dot{Y}_{2}  
\end{pmatrix}=\begin{pmatrix}
 0 & \omega_{m} & 0 & 0 & 0 & 0\\
  -\omega_{m} & -\gamma_{m} & G & g & G & g\\
   -g & 0 & -\kappa_1 & \Delta_{1}^{\prime} & 0 & -g_0 q_{s}\\
   G & 0 &-\Delta_{1}^{\prime} &-\kappa_1 & g_0 q_{s} & 0\\
   -g & 0 & 0 & -g_0 q_{s} & -\kappa_2 & \Delta_{1}^{\prime}\\
    G & 0 & g_0 q_{s} & 0 & -\Delta_{1}^{\prime} & -\kappa_2\\
\end{pmatrix} \begin{pmatrix}
\delta q \\ \delta p  \\ \delta X_{1} \\ \delta Y_{1} \\ \delta X_{2} \\ \delta Y_{2} 
\end{pmatrix} + \begin{pmatrix}
0 \\ \xi(t) \\ \sqrt{2\kappa_1}\delta X_{in} \\ \sqrt{2\kappa_1}\delta Y_{in} \\ \sqrt{2\kappa_2}\delta X_{in} \\ \sqrt{2\kappa_2}\delta Y_{in}
\end{pmatrix},
\end{equation}}
where $\Delta_{j}^{\prime}=\Delta_{j}-g_0 q_{s}, j=1,2.$, $G= \sqrt{2}g_0 Re[\alpha_{1}+\alpha_{2}]$, $g=\sqrt{2}g_0 Im[\alpha_{1}+\alpha_{2}]$. Here  
$ u=\begin{pmatrix}
\delta q, \delta p, \delta X_{1}, \delta Y_{1}, \delta X_{2}, \delta Y_{2}
\end{pmatrix}^T$ and $q_{s}, \alpha_{1}, \alpha_{2}$ are the steady state solutions of the mechanical and two cavity modes. $q_s=g_0\frac{|\alpha_1+\alpha_2|^2}{\omega_m}$. $\alpha_{1}$ can be solved from $-(i\Delta_1+\kappa_1)\alpha_1+i\frac{ g^2_0}{\omega_m} \alpha^3_1(1+\frac{i\Delta_1+\kappa_1}{i\Delta_2+\kappa_2})((1+\frac{\Delta_1 \Delta_2+\kappa_1 \kappa_2}{\Delta^2_2+\kappa^2_2})^2 +(\frac{\Delta_1 \kappa_2-\kappa_1 \Delta_2}{\Delta^2_2+\kappa^2_2})^2)+\eta_1=0$. $\alpha_2=\frac{i\Delta_1+\kappa_1}{i\Delta_2+\kappa_2}\alpha_1$.

\subsection{Coupling two Double-longitudinal-mode Cavity Optomechanical Systems by a BS}
In the frame rotating at the input field frequency ($\omega_{L}$), the quantum Langevin equations ($QLEs$) describing the system shown in Fig.6 can be written as: 
\begin{equation}
\begin{aligned} \label{eq4}
\dot{q}_{1}&=\omega_{m1}p_{1}\\
\dot{p}_{1}&=-\omega_{m1}q_{1}-\gamma_{m1}p_{1}+g_{0}(a_{1}+a_{2})^\dagger(a_{1}+a_{2})+\xi_{1}\\
\dot{a}_{1}&=-(i\Delta_{1}+\kappa_{1})a_{1}+ig_{0}(a_{1}+a_{2})q_{1}+\eta_{1}- \sqrt{2\kappa_{1}}a_{in}\\
\dot{a}_{2}&=-(i\Delta_{2}+\kappa_{2})a_{2}+ig_{0}(a_{1}+a_{2})q_{1}+\eta_{2}-\sqrt{2\kappa_{2}}a_{in}\\
\dot{q}_{2}&=\omega_{m2}p_{2}\\
\dot{p}_{2}&=-\omega_{m2}q_{2}-\gamma_{m2}p_{2}+g_{0}(a_{3}+a_{4})^\dagger(a_{3}+a_{4})+\xi_{2}\\
\dot{a}_{3}&=-(i\Delta_{3}+\kappa_{3})a_{3}+ig_{0}(a_{3}+a_{4})q_{2}+\eta_{3}-\sqrt{2\kappa_{3}}a_{in^{\prime}}\\
\dot{a}_{4}&=-(i\Delta_{4}+\kappa_{4})a_{4}+ig_{0}(a_{3}+a_{4})q_{2}+\eta_{4}-\sqrt{2\kappa_{4}}a_{in^{\prime}},
\end{aligned}
\end{equation}
where $\Delta_{j}=\omega_{j}-\omega_{L} (j=1, 2, 3, 4)$ is the frequency detuning of the cavity field $j$ with respect to the input laser field.
Then, the output $QLEs$ can be obtained, according to the input-output relationship ($a_{out}=\sqrt{2\kappa}a-a_{in}$). Afterwards, the steady-state $CM$ of the output modes can be achieved by solving the Lyapunov equation, and the final $CM$ can be written as:
 
\begin{equation}
\begin{aligned}
V=U\begin{pmatrix}V_{1} &0
\\0 &V_{2}\end{pmatrix} U^T,
\end{aligned}
\end{equation}

where $V_i (i=1,2)$ is the $CM$ of the $i$-th cavity and $U$ is the BS matrix, which can be described as\\

$U=\begin{pmatrix}
\mathbbm{1} &  &  &  &  & \\
 & cos\theta e^{i\phi} \mathbbm{1} & & & sin\theta e^{i(\phi+\pi)}\mathbbm{1} & \\
 &  & \mathbbm{1} &  &  &\\
 &  &  & \mathbbm{1} &  &  \\
 & sin\theta\mathbbm{1} &  &  & cos\theta\mathbbm{1} &  \\
 &  &  &  &  & \mathbbm{1}\\
\end{pmatrix}$,\\
where $\phi$ is the relative transmission phase angle of the cavity mode $1^{\prime}$ and $cos\theta$ is the transmission coefficient of the BS. $\mathbbm{1}$ is the 2$\times$2 identity matrix.

The matrices $\mathcal{A}_{1}$, $\mathcal{A}_{2}$, $\mathcal{D}_{1}$ and $\mathcal{D}_{2}$, based on which to get the CM  $V_i (i=1,2)$, are given by

$\mathcal{A}_{i}= \begin{pmatrix}
 0 & \omega_{mi} & 0 & 0 & 0 & 0\\
  -\omega_{mi} & -\gamma_{mi} & G^{Re}_{i} & G^{Im}_{i} & G^{Re}_{i} & G^{Im}_{i}\\
   -G^{Im}_{i} & 0 & -\kappa_{2i-1}+\sqrt{2\kappa_{2i-1}} & \Delta_{2i-1}-g_{0}q_{si} & 0 & -g_{0}q_{si}\\
   G^{Re}_{i} & 0 &-(\Delta_{2i-1}-g_{0}q_{si}) &-\kappa_{2i-1}+\sqrt{2\kappa_{2i-1}} & g_{0}q_{s1} & 0\\
   -G^{Im}_{i} & 0 & 0 & -g_{0}q_{si} & -\kappa_{2i}+\sqrt{2\kappa_{2i}} & \Delta_{2i}-g_{0}q_{si}\\
   G^{Re}_{i} & 0 & g_{0}q_{si} & 0 & -(\Delta_{2i}-g_{0}q_{si}) & -\kappa_{2i}+\sqrt{2\kappa_{2i}}\\
\end{pmatrix}$,

$\mathcal{D}_{i}=\begin{pmatrix}
 0 & 0 & 0 & 0 & 0 & 0 \\
0 & \gamma_{mi}(2{\overline{n}}+1) & 0 & 0 & 0 & 0\\
0 & 0 & \kappa_{2i-1} & 0 & \sqrt{\kappa_{2i-1}\kappa_{2i}} & 0\\
0 & 0 & 0 &\kappa_{2i-1} & 0 & \sqrt{\kappa_{2i-1}\kappa_{2i}} \\
0 & 0 & \sqrt{\kappa_{2i-1}\kappa_{2i}} & 0 & \kappa_{2i} & 0 \\
0 & 0 & 0 & \sqrt{\kappa_{2i-1}\kappa_{2i}} & 0 & \kappa_{2i} \\
\end{pmatrix}$, \\
where $G^{Re}_{i}= \sqrt{2}g_0 Re[\alpha_{2i-1}+\alpha_{2i}]$, $G^{Im}_{i}=\sqrt{2}g_0 Im[\alpha_{2i-1}+\alpha_{2i}]$, and $q_{s1}, q_{s2}, \alpha_{1}, \alpha_{2}, \alpha_{3}, \alpha_{4}$ are the steady state solutions of two mechanical and four cavity modes.
\subsection{Pumping the Double-longitudinal-mode Cavity Optomechanical System with two Orthogonal Polarizations}
In the frame rotating at the input field frequency $\omega_L$, the quantum Langevin equations ($QLEs$) describing the system that shown in Fig.10, can be written as:

{\begin{equation}\begin{aligned}
\dot{q} & = \omega_m p\\
\dot{p} & = -\omega_m q-\gamma_mp+g_0(a_1+a_2)^\dagger(a_1+a_2)+g_0(a_3+a_4)^\dagger(a_3+a_4)+\xi\\
\dot{a}_1 & = -(i\Delta_1+\kappa_1)a_1 +ig_0(a_1+a_2)q +\eta_1 +\sqrt{2\kappa_1}a_{in}\\
\dot{a}_2 & = -(i\Delta_2+\kappa_2)a_2 +ig_0(a_1+a_2)q +\eta_2 +\sqrt{2\kappa_2}a_{in}\\
\dot{a}_3 & = -(i\Delta_3+\kappa_3)a_3 +ig_0(a_3+a_4)q +\eta_3 +\sqrt{2\kappa_3}a_{in}\\
\dot{a}_4 & = -(i\Delta_4+\kappa_4)a_4 +ig_0(a_3+a_4)q +\eta_4 +\sqrt{2\kappa_4}a_{in},
\end{aligned}\end{equation}}
where $\Delta_j=\omega_j-\omega_L, (j=1, 2, 3, 4)$ is the frequency detuning of the cavity mode $j$ with respect to the input laser fields. Here the matrices $\mathcal{A}$ and $\mathcal{D}$ are given by

$\mathcal{A}= \begin{pmatrix}
 0 & \omega_{m} & 0 & 0 & 0 & 0 & 0 & 0 & 0 & 0\\
  -\omega_{m} & -\gamma_{m} & G & g & G & g & G & g & G & g\\
   -g & 0 & -\kappa_1 & \Delta_{1}^{\prime} & 0 & -g_{0}q_{s} & 0 & 0 & 0 & 0 \\
   G & 0 &-\Delta_{1}^{\prime} &-\kappa_1 & g_{0}q_{s} & 0 & 0 & 0 & 0 & 0\\
   -g & 0 & 0 & -g_{0}q_{s} & -\kappa_2 & \Delta_{2}^{\prime} & 0 & 0 & 0 & 0\\
    G & 0 & g_{0}q_{s} & 0 & -\Delta_{2}^{\prime} & -\kappa_2 & 0 & 0 & 0 & 0\\
-g & 0 & 0 & 0 & 0 & 0 & -\kappa_3 & \Delta_{3}^{\prime} & 0 & -g_{0}q_{s}\\
G & 0 & 0 & 0 & 0 & 0 & -\Delta_{3}^{\prime} & -\kappa_3 & g_{0}q_{s} & 0\\
-g & 0 & 0 & 0 & 0 & 0 & 0 & -g_{0}q_{s} & -\kappa_4 & \Delta_{4}^{\prime} \\
G & 0 & 0 & 0 & 0 & 0 & g_{0}q_{s} & 0 & -\Delta_{4}^{\prime} & -\kappa_4\\
\end{pmatrix}$,\\

$\mathcal{D}=\begin{pmatrix}
0 & 0 & 0 & 0 & 0 & 0 & 0 & 0 & 0 & 0 \\
0 & \gamma_{m}(2{\overline{n}}+1) & 0 & 0 & 0 & 0 & 0 & 0 & 0 & 0 \\
0 & 0 & \kappa_1 & 0 & \sqrt{\kappa_1\kappa_2} & 0 & \sqrt{\kappa_1\kappa_3} & 0 & \sqrt{\kappa_1\kappa_4} & 0 \\
0 & 0 & 0 & \kappa_1 & 0 & \sqrt{\kappa_1\kappa_2} & 0 & \sqrt{\kappa_1\kappa_3} & 0 & \sqrt{\kappa_1\kappa_4} \\
0 & 0 & \sqrt{\kappa_1\kappa_2} & 0 & \kappa_2 & 0 & \sqrt{\kappa_2\kappa_3} & 0 & \sqrt{\kappa_2\kappa_4} & 0 \\
0 & 0 & 0 & \sqrt{\kappa_1\kappa_2} & 0 & \kappa_2 & 0 & \sqrt{\kappa_2\kappa_3} & 0 & \sqrt{\kappa_2\kappa_4} \\ 0 & 0 & \sqrt{\kappa_1\kappa_3} & 0 & \sqrt{\kappa_2\kappa_3} & 0 & \kappa_3 & 0 & \sqrt{\kappa_3\kappa_4} & 0 \\
0 & 0 & 0 & \sqrt{\kappa_1\kappa_3} & 0 & \sqrt{\kappa_2\kappa_3} & 0 & \kappa_3 & 0 & \sqrt{\kappa_3\kappa_4} \\
0 & 0 & \sqrt{\kappa_1\kappa_4} & 0 & \sqrt{\kappa_2\kappa_4} & 0 & \sqrt{\kappa_3\kappa_4} & 0 & \kappa_4 & 0 \\
0 & 0 & 0 & \sqrt{\kappa_1\kappa_4} & 0 & \sqrt{\kappa_2\kappa_4} & 0 & \sqrt{\kappa_3\kappa_4} & 0 & \kappa_4 \\
\end{pmatrix}$,
 where $G= \sqrt{2}g_{0}Re[\alpha_{1}+\alpha_{2}]=\sqrt{2}g_{0}Re[\alpha_{3}+\alpha_{4}]$, $g=\sqrt{2} g_0 Im[\alpha_{1}+\alpha_{2}]=\sqrt{2} g_0 Im[\alpha_{3}+\alpha_{4}]$  $\Delta_{j}^{\prime} =\Delta_{j}-g_{0}q_{s} $ and $q_{s}, \alpha_{1}, \alpha_{2}, \alpha_{3}, \alpha_{4}$ are the steady state solutions of the mechanical and four cavity modes.
%%%%%%%%% References %%%%%%%%%%%%
\bibliography{ref}

\begin{thebibliography}{10}
\newcommand{\enquote}[1]{``#1''}

\bibitem{169}
S.~Pirandola and S.~L. Braunstein, \enquote{Physics: Unite to build a quantum
  internet,} {\protect\JournalTitle{Nature}} \textbf{532}, 169--171 (2016).

\bibitem{Guo2019}
Y.~Guo, B.-H. Liu, C.-F. Li, and G.~Guo, \enquote{Advances in quantum dense
  coding,} {\protect\JournalTitle{Advanced Quantum Technologies}} \textbf{2},
  1900011 (2019).

\bibitem{45}
T.~D. Ladd, F.~Jelezko, R.~Laflamme, Y.~Nakamura, C.~R. Monroe, and J.~L.
  O'Brien, \enquote{Quantum computers,} {\protect\JournalTitle{Nature}}
  \textbf{464}, 45--53 (2010).

\bibitem{965}
N.~Gisin and R.~Thew, \enquote{Quantum communication technology,}
  {\protect\JournalTitle{Electronics Letters}} \textbf{46}, 965--967 (2010).

\bibitem{01226}
K.~Fang, J.~Zhao, X.~Li, Y.~Li, and R.~Duan, \enquote{Quantum network: from
  theory to practice,} {\protect\JournalTitle{ArXiv}} \textbf{2212. 01226}
  (2022).

\bibitem{073601}
F.~A.~S. Barbosa, A.~S. Coelho, L.~F. Mu{\~n}oz-Mart{\'i}nez,
  L.~Ortiz-Guti{\'e}rrez, A.~S. Villar, P.~Nussenzveig, and M.~Martinelli,
  \enquote{Hexapartite entanglement in an above-threshold optical parametric
  oscillator.} {\protect\JournalTitle{Phys. Rev. Lett.}} \textbf{121 7}, 073601
  (2017).

\bibitem{167}
S.~C. Armstrong, M.~Wang, R.~Y. Teh, Q.~Gong, Q.~He, J.~Janousek, H.-A. Bachor,
  M.~D. Reid, and P.~K. Lam, \enquote{Multipartite einstein–podolsky–rosen
  steering and genuine tripartite entanglement with optical networks,}
  {\protect\JournalTitle{Nature Physics}} \textbf{11}, 167 -- 172 (2014).

\bibitem{96}
H.~Tan, Y.~hua Wei, and G.~xiang Li, \enquote{Building mechanical
  greenberger-horne-zeilinger and cluster states by harnessing optomechanical
  quantum steerable correlations,} {\protect\JournalTitle{Physical Review A}}
  \textbf{96} (2017).

\bibitem{124205}
J.~Zhang, Y.~Wang, X.-Y. Liu, and R.~Yang, \enquote{Parallel generation of 31
  tripartite entangled states based on optical frequency combs,}
  {\protect\JournalTitle{Chinese Physics B}} \textbf{26}, 124205 (2017).

\bibitem{7535}
P.~Du, Y.~Wang, K.~Liu, R.~Yang, and J.~Zhang, \enquote{Generation of
  large-scale continuous-variable cluster states multiplexed both in time and
  frequency domains,} {\protect\JournalTitle{Optics express}} \textbf{31 5},
  7535--7544 (2023).

\bibitem{1098}
Z.-S. Yuan, Y.-A. Chen, B.~Zhao, S.~Chen, J.~Schmiedmayer, and J.-W. Pan,
  \enquote{Experimental demonstration of a bdcz quantum repeater node,}
  {\protect\JournalTitle{Nature}} \textbf{454}, 1098--1101 (2008).

\bibitem{093601}
W.~Qin, A.~Miranowicz, P.-B. Li, X.-Y. L\"u, J.~Q. You, and F.~Nori,
  \enquote{Exponentially enhanced light-matter interaction, cooperativities,
  and steady-state entanglement using parametric amplification,}
  {\protect\JournalTitle{Phys. Rev. Lett.}} \textbf{120}, 093601 (2018).

\bibitem{153}
B.~B. Blinov, D.~L. Moehring, L.~Duan, and C.~R. Monroe, \enquote{Observation
  of entanglement between a single trapped atom and a single photon,}
  {\protect\JournalTitle{Nature}} \textbf{428}, 153--157 (2004).

\bibitem{281}
D.~Leibfried, R.~Blatt, C.~Monroe, and D.~Wineland, \enquote{Quantum dynamics
  of single trapped ions,} {\protect\JournalTitle{Rev. Mod. Phys.}}
  \textbf{75}, 281--324 (2003).

\bibitem{426}
W.~Gao, P.~Fallahi, E.~Togan, J.~Miguel-S{\'a}nchez, and A.~Imamoğlu,
  \enquote{Observation of entanglement between a quantum dot spin and a single
  photon,} {\protect\JournalTitle{Nature}} \textbf{491}, 426--430 (2012).

\bibitem{20171}
X.~Gu, A.~F. Kockum, A.~Miranowicz, Y.~xi~Liu, and F.~Nori, \enquote{Microwave
  photonics with superconducting quantum circuits,}
  {\protect\JournalTitle{Physics Reports}} \textbf{718-719}, 1--102 (2017).

\bibitem{589}
J.~Q. You and F.~Nori, \enquote{Atomic physics and quantum optics using
  superconducting circuits,} {\protect\JournalTitle{Nature}} \textbf{474},
  589--597 (2011).

\bibitem{042330}
Q.~Cai, J.~Liao, and Q.~Zhou, \enquote{Entangling two microwave modes via
  optomechanics,} {\protect\JournalTitle{Phys. Rev. A}} \textbf{100}, 042330
  (2019).

\bibitem{17237}
Z.~X. Chen, Q.~Lin, B.~He, and Z.~Y. Lin, \enquote{Entanglement dynamics in
  double-cavity optomechanical systems.} {\protect\JournalTitle{Optics
  express}} \textbf{25 15}, 17237--17248 (2017).

\bibitem{042342}
S.~Barzanjeh, D.~Vitali, P.~Tombesi, and G.~J. Milburn, \enquote{Entangling
  optical and microwave cavity modes by means of a nanomechanical resonator,}
  {\protect\JournalTitle{Phys. Rev. A}} \textbf{84}, 042342 (2011).

\bibitem{054061}
C.~Zhong, X.~Han, and L.~Jiang, \enquote{Microwave and optical entanglement for
  quantum transduction with electro-optomechanics,}
  {\protect\JournalTitle{Phys. Rev. Appl.}} \textbf{18}, 054061 (2022).

\bibitem{29581}
T.~Wang, L.~Wang, Y.~Liu, C.-H. Bai, D.-Y. Wang, H.~Wang, and S.~Zhang,
  \enquote{Temperature-resistant generation of robust entanglement with
  blue-detuning driving and mechanical gain.} {\protect\JournalTitle{Optics
  express}} \textbf{27 21}, 29581--29593 (2019).

\bibitem{063602}
D.-G. Lai, J.-Q. Liao, A.~Miranowicz, and F.~Nori, \enquote{Noise-tolerant
  optomechanical entanglement via synthetic magnetism,}
  {\protect\JournalTitle{Phys. Rev. Lett.}} \textbf{129}, 063602 (2022).

\bibitem{0485}
Y.~Li, Y.-F. Jiao, J.-X. Liu, A.~Miranowicz, Y.-L. Zuo, L.-M. Kuang, and
  H.~Jing, \enquote{Vector optomechanical entanglement,}
  {\protect\JournalTitle{Nanophotonics}} \textbf{11}, 67--77 (2022).

\bibitem{559}
V.~Giovannetti, S.~Mancini, and P.~Tombesi, \enquote{Radiation pressure induced
  einstein-podolsky-rosen paradox,} {\protect\JournalTitle{Europhysics
  Letters}} \textbf{54}, 559 (2001).

\bibitem{053515}
X.~Z. Hao, X.~Y. Zhang, Y.~H. Zhou, W.~Li, S.~C. Hou, and X.~X. Yi,
  \enquote{Dynamical bipartite and tripartite entanglement of mechanical
  oscillators in an optomechanical array,} {\protect\JournalTitle{Phys. Rev.
  A}} \textbf{104}, 053515 (2021).

\bibitem{15032}
R.-X. Chen, C.-G. Liao, and X.-M. Lin, \enquote{Dissipative generation of
  significant amount of mechanical entanglement in a coupled optomechanical
  system,} {\protect\JournalTitle{Scientific Reports}} \textbf{7} (2017).

\bibitem{063801}
H.~Miao, C.~Zhao, L.~Ju, and D.~G. Blair, \enquote{Quantum ground-state cooling
  and tripartite entanglement with three-mode optoacoustic interactions,}
  {\protect\JournalTitle{Phys. Rev. A}} \textbf{79}, 063801 (2009).

\bibitem{1}
C.~Zhao, Q.~Fang, S.~Susmithan, H.~Miao, L.~Ju, Y.~Fan, D.~G. Blair, D.~J.
  Hosken, J.~Munch, P.~J. Veitch, and B.~J.~J. Slagmolen,
  \enquote{High-sensitivity three-mode optomechanical transducer,}
  {\protect\JournalTitle{Physical Review A}} \textbf{84}, 1--6 (2011).

\bibitem{485}
Y.-H. Ma and X.-F. Zhang, \enquote{Genuine quadripartite macroscopic
  entanglement generated in two-mode optomechanical systems,}
  {\protect\JournalTitle{Applied Physics B}} \textbf{112}, 485--489 (2013).

\bibitem{052303}
X.~Yang, Y.~Ling, X.~Shao, and M.~Xiao, \enquote{Generation of robust
  tripartite entanglement with a single-cavity optomechanical system,}
  {\protect\JournalTitle{Physical Review A}} \textbf{95}, 052303 (2017).

\bibitem{033842}
Z.~J. Deng, X.-B. Yan, Y.-D. Wang, and C.-W. Wu, \enquote{Optimizing the
  output-photon entanglement in multimode optomechanical systems,}
  {\protect\JournalTitle{Phys. Rev. A}} \textbf{93}, 033842 (2016).

\bibitem{10306}
C.-G. Liao, X.~Shang, H.-Y. Xie, and X.~Lin, \enquote{Dissipation-driven
  entanglement between two microwave fields in a four-mode hybrid cavity
  optomechanical system.} {\protect\JournalTitle{Optics express}} \textbf{30
  7}, 10306--10316 (2022).

\bibitem{042320}
C.~Jiang, S.~Tserkis, K.~Collins, S.~Onoe, Y.~Li, and L.~Tian,
  \enquote{Switchable bipartite and genuine tripartite entanglement via an
  optoelectromechanical interface,} {\protect\JournalTitle{Phys. Rev. A}}
  \textbf{101}, 042320 (2020).

\bibitem{1244563}
T.~A. Palomaki, J.~D. Teufel, R.~W. Simmonds, and K.~W. Lehnert,
  \enquote{Entangling mechanical motion with microwave fields,}
  {\protect\JournalTitle{Science}} \textbf{342}, 710 -- 713 (2013).

\bibitem{651}
S.~T. Velez, V.~Sudhir, N.~Sangouard, and C.~Galland, \enquote{Bell
  correlations between light and vibration at ambient conditions.}
  {\protect\JournalTitle{Science advances}} \textbf{6 51} (2020).

\bibitem{478}
C.~F. Ockeloen-Korppi, E.~Damsk{\"a}gg, J.-M. Pirkkalainen, M.~Asjad, A.~A.
  Clerk, F.~Massel, M.~J. Woolley, and M.~A. Sillanp{\"a}{\"a},
  \enquote{Stabilized entanglement of massive mechanical oscillators,}
  {\protect\JournalTitle{Nature}} \textbf{556}, 478--482 (2018).

\bibitem{473}
R.~Riedinger, A.~Wallucks, I.~Marinkovi{\'c}, C.~L{\"o}schnauer, M.~Aspelmeyer,
  S.~Hong, and S.~Gr{\"o}blacher, \enquote{Remote quantum entanglement between
  two micromechanical oscillators,} {\protect\JournalTitle{Nature}}
  \textbf{556}, 473--477 (2017).

\bibitem{1391}
M.~Aspelmeyer, T.~J. Kippenberg, and F.~Marquardt, \enquote{Cavity
  optomechanics,} {\protect\JournalTitle{Rev. Mod. Phys.}} \textbf{86},
  1391--1452 (2014).

\bibitem{113601}
K.~Zhang, F.~Bariani, Y.~Dong, W.~Zhang, and P.~Meystre, \enquote{Proposal for
  an optomechanical microwave sensor at the subphoton level,}
  {\protect\JournalTitle{Phys. Rev. Lett.}} \textbf{114}, 113601 (2015).

\bibitem{120602}
A.~Mari and J.~Eisert, \enquote{Cooling by heating: Very hot thermal light can
  significantly cool quantum systems,} {\protect\JournalTitle{Phys. Rev.
  Lett.}} \textbf{108}, 120602 (2012).

\bibitem{023812}
H.~K. Cheung and C.~K. Law, \enquote{Nonadiabatic optomechanical hamiltonian of
  a moving dielectric membrane in a cavity,} {\protect\JournalTitle{Phys. Rev.
  A}} \textbf{84}, 023812 (2011).

\bibitem{032314}
G.~Vidal and R.~F. Werner, \enquote{Computable measure of entanglement,}
  {\protect\JournalTitle{Phys. Rev. A}} \textbf{65}, 032314 (2002).

\end{thebibliography}

\end{document}